\documentclass[twocolumn,showpacs,preprintnumbers,amsmath,amssymb,prb,floatfix]{revtex4}

\usepackage{graphicx}
\usepackage{dcolumn}
\usepackage{bm}
\usepackage{epstopdf}

\begin{document}
\bibliographystyle{prsty}

\title{Electronic and Magnetic Structure of the (LaMnO$_3$)$_{2n}$/(SrMnO$_3$)$_n$ Superlattices}

\author{B. R. K. Nanda}
\author{S. Satpathy}%
\affiliation{%
Department of Physics and Astronomy, University of Missouri,
Columbia, MO 65211}%

\date{\today}

\begin{abstract}

We study the magnetic structure of the (LaMnO$_3$)$_{2n}$/(SrMnO$_3$)$_n$ superlattices from density-functional calculations. In agreement with the experiments, we find that the magnetism changes with the layer thickness `n'. The reason for the different magnetic structures is shown to be the varying  potential barrier across the interface, which controls the leakage of the Mn-e$_g$ electrons from the LMO side to the SMO side. This in turn affects the 
interfacial magnetism via the carrier-mediated Zener double exchange. For n=1 superlattice, the Mn-e$_g$ electrons are more or less spread over the entire lattice, so that the magnetic behavior is similar to the equivalent alloy compound La$_{2/3}$Sr$_{1/3}$MnO$_3$. For larger n, the e$_g$ electron transfer occurs mostly between the two layers adjacent to the interface,  thus leaving the magnetism unchanged and bulk-like away from the interface region.
\end{abstract}

\pacs{75.70.Cn, 71.20.-b, 73.20.-r}
 
\maketitle

{\it Introduction --} Superlattices made up of strongly correlated transition metal oxides such as LaMnO$_3$ (LMO) and SrMnO$_3$ (SMO) are of current interest because of the diverse magnetic and electronic phases they exhibit. For example recent experimental results reveal that (LMO)$_{2n}$/(SMO)$_n$ superlattice is uniformly ferromagnetic for the short period structure (n = 1) while the long period superlattices (n $\ge$ 3) show bulk antiferromagnetic ordering away from the interface and ferromagnetic ordering at the interface\cite{anand,adamo}.

   In this paper, we report results of our electronic structure calculations, based on the density functional theory (DFT), performed to understand the change 
in the magnetic properties of the (LMO)$_{2n}$/(SMO)$_n$ superlattices as a function of the layer thickness `n'. We show that there exists a potential barrier for the electrons, in particular, for the 
Mn-e$_g$ electrons, the strength of which differs with the layer thickness n. This varying potential barrier, which 
controls the leakage of the Mn-e$_g$ electrons from the LMO side to SMO side, in turn determines the stable magnetic 
configurations in the (LMO)$_{2n}$/(SMO)$_n$ superlattices. In agreement with the experiments, our calculations 
predict a uniform ferromagnetic (FM) ordering in the short period superlattice (n = 1) and the co-existence of interface FM phase and inner bulk antiferromagnetic (AFM) phases in the long period superlattices (n $\ge$ 3). The magnetism can be qualitatively understood in terms of the two competing interactions, viz., the
antiferromagnetic superexchange between the core spins and the 
Zener ferromagnetic double exchange mediated by the itinerant e$_g$ electrons.

\begin{figure}
\includegraphics[width=5.5cm]{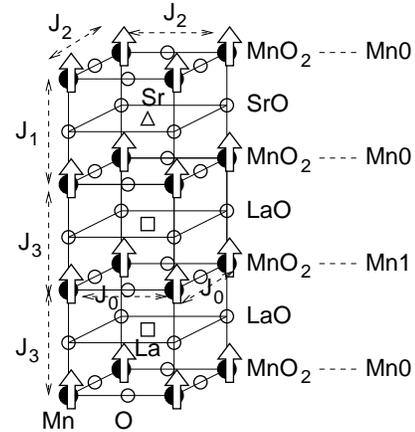}
\caption{\label{l2s1fig} Schematic unit cell of (LMO)$_2$/(SMO)$_1$ superlattice and the magnetic structure as predicted from the DFT calculations. Mn0 represents the interfacial Mn atoms surrounded by both SrO and LaO layers and Mn1 represents the Mn atoms inside the LMO part. Because the SMO part is small, there is no Mn atom surrounded by two SrO layers in this structure. The  nearest neighbor Mn - Mn exchange interactions are indicated by the J's. }
\end{figure}

\vspace{0.4cm}
{\it Computational and structural details --} The results presented in this paper are obtained from the DFT studies of three superlattices, namely, 
(LMO)$_2$/(SMO)$_1$ (schematically shown in Fig. \ref{l2s1fig}), (LMO)$_4$/(SMO)$_2$, and (LMO)$_6$/(SMO)$_3$ using the 
linear muffin-tin orbitals (LMTO) method\cite{lmto} with general gradient approximation and 
on-site Coulomb correction (GGA+U)\cite{gga,ucorrec}. The Coulomb (U) and the exchange parameter (J) are taken as 5 eV and 1 eV respectively.  Each superlattice consists of twice the formula unit because of the 
magnetic structures considered in the paper. 

The bulk lattice parameters of LMO and SMO are respectively 3.935 {\AA} and 3.802 {\AA}. However, since most of the 
experimental results reported in the literature are based on the LMO/SMO superlattices grown on the SrTiO$_3$ (STO)
 substrate\cite{anand, adamo,satoh,may}, we have taken the in-plane lattice parameter for the (LMO)$_{2n}$/(SMO)$_n$ superlattices 
as the bulk STO lattice parameter (3.905 {\AA}). 
The out-of-plane lattice parameters are taken to be 3.99 {\AA} (LMO) and 3.65 {\AA} (SMO) in order to preserve the bulk volumes.
The basal John-Teller distortion (Q$_2$)for the inner Mn layers in the LMO site is taken the same as the bulk value (0.15 {\AA}). The value of Q$_2$ for the interface Mn layers is taken as 0.07 {\AA} in view of the fact that the JT distortion is reduced in the mixed compound (La, Sr)MnO$_3$. By growing the superlattices on different substrates, the strain condition of the structure can be changed, which then would affect magnetism via the change in the electron hopping due to different orbital ordering. The effect of strain on the magnetism has been discussed in detail elsewhere.\cite{Ranjit-strain}

\begin{figure}
\includegraphics[width=5.6cm]{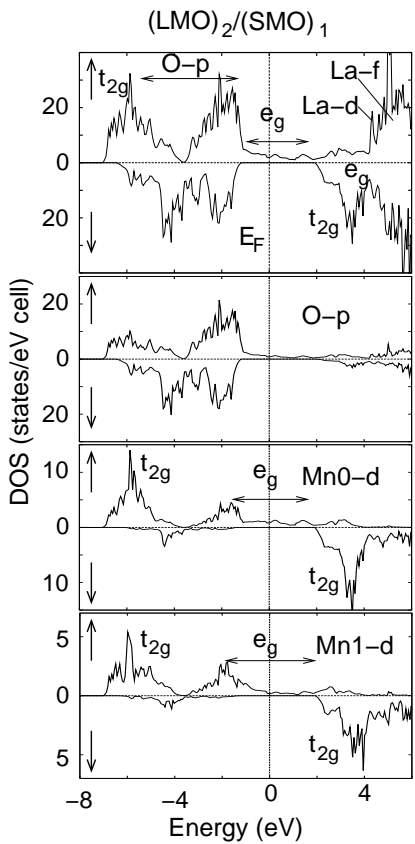}
\caption{\label{dosl2s1} Total (upper panel) and partial spin-resolved DOS for the ferromagnetic (LMO)$_2$/(SMO)$_1$ superlattice. The labeling of the Mn atoms are as in Fig. \ref{l2s1fig}. Upper  and lower segments within each panel correspond, respectively, to the majority ($\uparrow$)  and minority ($\downarrow$) spin densities.}
\end{figure}

\vspace{0.4cm}
{\it Electronic structure of the (LMO)$_2$/(SMO)$_1$ superlattice --} Before discussing the electronic and magnetic properties of the (LMO)$_{2n}$/(SMO)$_n$ superlattices, we summarize
 the electronic structure and magnetism of the bulk SMO and LMO compounds. In bulk SMO, the Mn atoms are in the 4+ charged state, so that they have three d-electrons occupying the triply degenerate t$_{2g}$ states. The doubly degenerate 
e$_g$ states, which are higher in energy with respect to the t$_{2g}$ states because of the MnO$_6$ octahedral crystal 
field, remain unoccupied. The t$^{3}_{2g}$ core spins interact via an antiferromagnetic superexchange 
so as to stabilize the G-type AFM ordering in the bulk SMO compound\cite{millisgafm,zhougafm}. 

 In bulk LMO, the Mn atoms are in the 3+ charged state with four occupied d-electrons. Three electrons are present in the t$_{2g}$ states and the remaining one in the e$_g$ states. The Jahn-Teller distortion 
of the MnO$_6$ octahedron further splits the e$_g$ states into two non-degenerate states: e$_{g}^{1}$ which is lower in 
energy and e$_{g}^{2}$ which is higher in energy\cite{sashilmo}. The  e$_{g}^{1}$ 
orbital, occupied by the lone electron, has its lobes pointed towards the longest Mn-O bond. The JT distortion stabilizes the A-type AFM
structure in the LMO compound due to a combination of the superexchange and Zener double exchange \cite{goodenough}. The charge reconstruction at the LMO/SMO interface\cite{koida,smadici} is expected to change the electronic and magnetic properties of the (LMO)$_{2n}$/(SMO)$_n$ superlattices which will be discussed in the remaining part of the paper.

 Out of a number of magnetic configurations that we considered, the DFT calculations predict a ferromagnetic ground state for the (LMO)$_2$/(SMO)$_1$ superlattice. In Fig. \ref{dosl2s1}, we have shown the total and partial spin-resolved densities-of-states (DOS) for the ferromagnetic configuration of this superlattice. The characteristic features of the electronic structure as seen from the figure are as follows. The Mn-t$_{2g}$ states lie far below the Fermi level (E$_F$) because of the octahedral crystal field and strong Coulomb repulsion. The O-p states occur in the energy range of -6 to -1 eV. The Mn-e$_g$ states occur around the Fermi level E$_F$, while the Sr-d, La-d and La-f states lie far above it.

 As Fig. \ref{dosl2s1} shows, the most important feature in the electronic structure of (LMO)$_2$/(SMO)$_1$ is that the delocalized 
e$_g$ states of both Mn0 (Mn atoms at the interface) and Mn1 (Mn atoms inside the LMO part) are partially occupied which 
is in agreement with the earlier electronic structure calculations\cite{picozzi}. These partially occupied e$_g$ states 
will mediate a strong Zener ferromagnetic double exchange\cite{zener,anderson,gennes} between the Mn-t$_{2g}$ core spins 
which wins over the antiferromagnetic superexchange, so that  a uniform ferromagnetic ordering throughout the superlattice is
stabilized. The calculation of the Mn - Mn exchange interactions discussed below indicates that the FM ordering is stable, quite similar to the equivalent alloy compound La$_{2/3}$Sr$_{1/3}$MnO$_3$\cite{anand}.

\begin{figure}
\includegraphics[width=7.5cm]{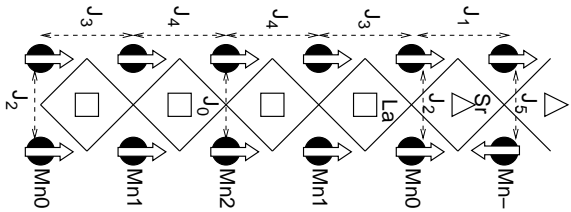}
\caption{\label{l4s2fig} Schematic unit cell of (LMO)$_4$/(SMO)$_2$ superlattice and the magnetic structure as obtained from the DFT calculations. Oxygen atoms occur at the intersections of the checkered lines forming the MnO$_6$ octahedron. Mn atoms of each MnO$_2$ layer are labeled as shown in the figure. Definitions of the exchange interactions for the (LMO)$_6$/(SMO)$_3$ superlattice are identical to the ones shown here and they are also consistent with Fig. \ref{l2s1fig} for the (LMO)$_4$/(SMO)$_2$ superlattice.}
\end{figure}

\vspace{0.4cm}
{\it Magnetic exchange interaction --} In order to study the magnetic ground state for the (LMO)$_{2n}$/(SMO)$_n$ superlattices, we have calculated the neighboring Mn - Mn exchange interaction energies (J's) for various exchange interactions as shown in Figs. \ref{l2s1fig} and \ref{l4s2fig}. In these 
figures the symbol J$_1$ represents the out-of-plane exchange interactions across the SrO layer close to the interface,
 while J$_3$ and J$_4$ represent the same across the LaO layers close to the interface and away from the interface 
respectively. J$_2$ denotes the in-plane exchange interaction for the interfacial MnO$_2$ layer which is surrounded by LaO and SrO layers, whereas J$_0$ and J$_5$ denote the same for the MnO$_2$ layer inside the LMO and SMO part of the superlattice respectively. 

The exchange interactions are calculated by taking the energy difference between the ferromagnetic alignment and the
antiferromagnetic alignment of two neighboring Mn spins (J = E$_{\uparrow\uparrow}$ - E$_{\uparrow\downarrow}$). 
The results are listed in Table I. The in-plane magnetic interaction J$_0$ inside the LMO part, which is strongly ferromagnetic, was not computed. For the case of the (LMO)$_6$/(SMO)$_3$ superlattice, the values of the exchange interaction for the Mn layers away from the interface in the LMO part and SMO part are respectively 12 meV (J$_4$) and 19 meV (J$_5$). These values are in good agreement with the experimental results for the bulk LMO (J $\sim$ 9.7 meV) and bulk SMO (J $\sim$ 13.1 meV)\cite{hakim,hirota,wollan}.

 From Table I, we see that for (LMO)$_2$/(SMO)$_1$, the in-plane exchange interaction J$_1$ as well as
 the out-of-plane exchange interactions J$_2$ and J$_3$ are strong and negative so as to stabilize the FM ordering throughout the superlattice consistent with the experimental observations\cite{anand,adamo}.   
Turning now to the (LMO)$_4$/(SMO)$_2$ superlattice,  the in-plane interactions (J$_0$ and J$_2$) are FM as also are the out-of-plane interactions ( J$_3$ and J$_4$ ) within the LMO part. 
In the SMO part, the out-of-plane exchange interaction (J$_1$) is AFM, but this being weaker as compared to the in-plane J$_2$ (FM) and J$_5$ (AFM), the magnetic configuration within the SMO part is controlled by the latter two exchange interactions as shown in Fig. 3.
 The values of J's  in the (LMO)$_6$/(SMO)$_3$ superlattice are similar to those of the (LMO)$_4$/(SMO)$_2$ superlattice, except that now the out-of-plane exchange interaction for the inner MnO$_2$ layers in the LMO side (J$_4$) is positive so as to establish an A-type AFM configuration as in the bulk LMO compound.  

\begin{table}
\begin{center}
\begin{tabular}{c|ccccc}
\hline
Superlattice&J$_1$&J$_2$&J$_3$&J$_4$&J$_5$\\
\hline
(LMO)$_2$/(SMO)$_1$&-11&-39&-26&&\\
(LMO)$_4$/(SMO)$_2$&10&-36&-18&-4&17\\
(LMO)$_6$/(SMO)$_3$&14&-37&-6&12&19\\
\hline
\end{tabular}
\caption{Calculated magnetic exchange interactions in units of meV. The exchange energies are obtained from the energy difference between the FM and the AFM alignments of two neighboring Mn spins. Negative (positive) value of the exchange energy indicates the FM (AFM) ordering.
 }
\end{center}
\end{table}
\begin{figure}[htbp]
\includegraphics[width=7.5cm]{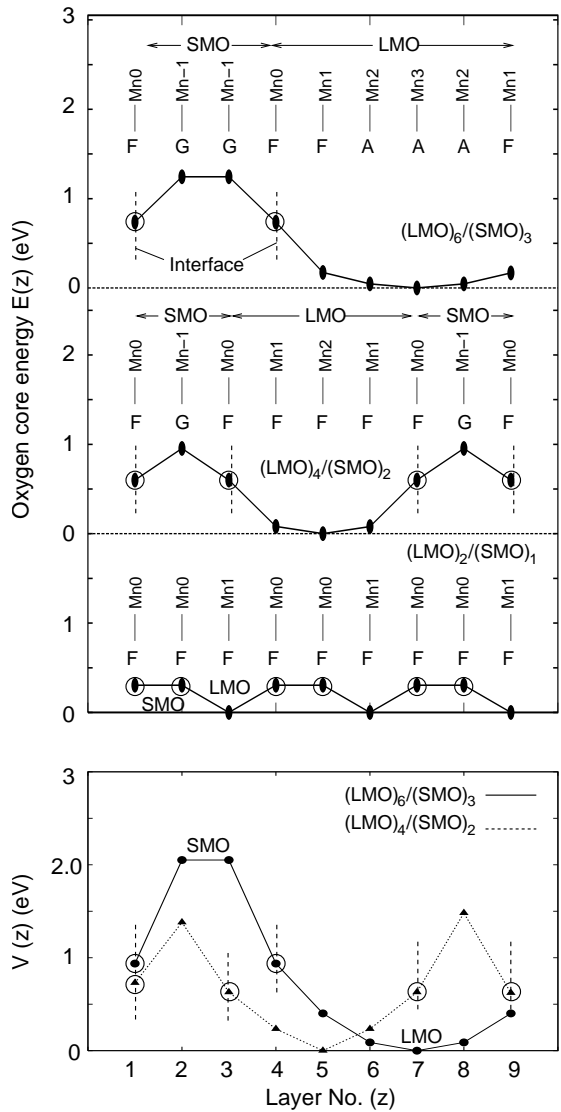}
\caption{\label{potential} Variations of the oxygen 1s core energy (upper panel) and the energy of the lowest Mn-e$_g$ state (lower panel) of each MnO$_2$ layer, obtained from the layer-projected wave function characters. Mn atoms of each MnO$_2$ layer are labeled as shown in the figure. The interfacial manganese atoms (Mn0), which are sandwiched by the LaO and SrO layers, are shown by open circles with vertical dashed lines, indicating the position of the interface. The magnetic ordering of Mn spins for each layer as obtained from the DFT calculations is shown with the symbols F (FM), G (G-AFM) and A (A-AFM).
A potential barrier is clearly seen for the n=2 and n=3 superlattices. }
\end{figure}

We note from the above discussions that as we increase the layer thickness `n', the FM interactions between the Mn spins occurring on the two sides of the LaO layers ( see J$_3$ and J$_4$ in Table I) gradually become weak, which eventually makes the LMO part type-A AFM like in the bulk. This already happens for n=3. The transition from the FM to AFM ordering for the Mn layers away from the interface with the increase of the layer thickness `n' is indicative of the fact that the charge reconstruction is essentially confined to the few interface layers for the long period superlattices (n $\ge$ 3).

%
\vspace{0.4cm}
{\it Electric potential profile and charge reconstruction at the interface --}
The potential seen by the electrons varies as one crosses the interface from one side to the other.
This for example leads to the well-known band offset in the semiconductors. Our calculations show that
for the present superlattices, there is a potential barrier as one goes from the LMO to the SMO side. This controls the 
leakage of the Mn-e$_g$ electrons across the barrier, which in turn affects the magnetic exchange interactions near the 
interface leading to diverse magnetic phases. 

In Fig. \ref{potential} (upper panel), we plot the calculated  oxygen 1s core energies indicating the potential barrier across the interface. However, the valence states experience a somewhat different potential 
than the core states because of different energy terms. Since the Mn-e$_g$ electrons are mainly the electrons that are 
transferred across the interface, we now examine the potential V(z) felt by these electrons. In order to obtain the variation of this potential, we have studied the 
band structure and the atomic characters of the wave functions in each superlattice by examining the so-called  `fat' bands in the 
LMTO results, which indicate the relative contributions of the various orbitals to the wave function making the band. From the `fat' bands, the lowest Mn-e$_g$ state belonging to a particular Mn layer can be identified, which is then 
indicative of the potential experienced by the Mn-e$_g$ electrons in the various layers. 

These results are shown in Fig. \ref{potential} (lower panel). The variation of V(z) for the n =1 superlattice is quite similar to the variation of the oxygen 1s core energies and hence is not shown in the figure.
For this superlattice, we have a weakly varying potential due to the close proximity of the interfaces to one another, which results in the overlap of the attractive Coulomb potential formed by the positively charged interfacial (LaO)$^+$ layers. In this case the Mn-e$_g$ electrons are more or less spread throughout the superlattice as seen from the layer projected DOS (Fig. \ref{dosl2s1}), where all Mn atoms have partially filled e$_g$ states. These itinerant e$_g$ electrons mediate the Zener double exchange stabilizing the FM ordering  throughout the superlattice. 

\begin{figure}
\includegraphics[width=6.0cm]{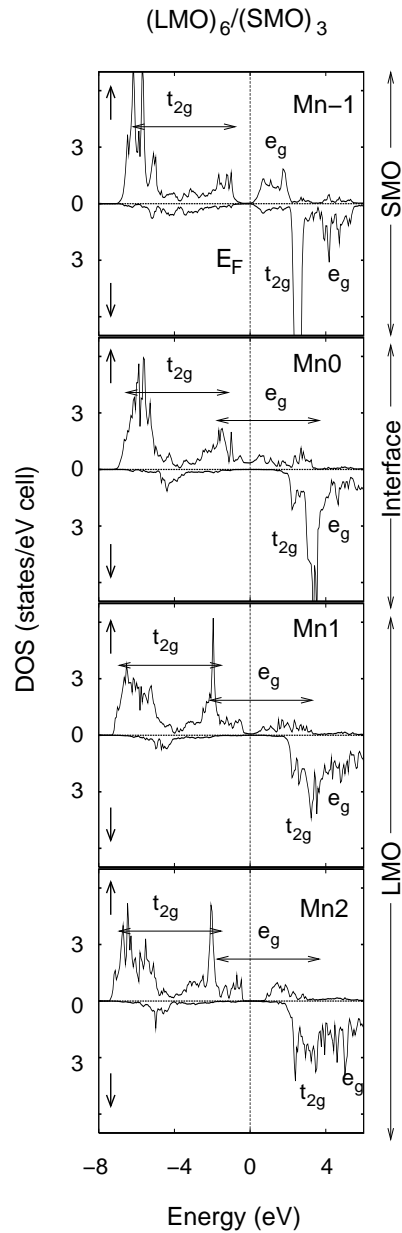} 
\caption{\label{dosl6s3} Spin-up ($\uparrow$) and spin-down ($\downarrow$) Mn-d DOS for the n = 3 superlattice. Up and down spins are with respect to the local magnetic moment of the Mn atom. The labeling of the Mn atoms are as in Fig. \ref{potential}. The projected Mn3 densities (not shown here) are similar to the Mn2 densities
 as the bulk limit has already been reached.}
\end{figure}

   With the increase of the layer thickness `n,' the variation of the potential becomes stronger leading to the formation 
of a potential barrier at the interface with the LMO side having a lower potential than the SMO side. This results in restricting the leakage of the Mn-e$_g$ electrons to the SMO side (Fig. 4). Thus, for example in the case of the n = 3 superlattice, there is very little e$_g$ electron on the Mn-1 atom belonging to the SMO side (Fig. \ref{dosl6s3}, topmost panel). Since the Mn-e$_g$ states are unoccupied in the SMO side, a G-type AFM structure is stabilized as in the bulk SMO.
    
   The case of the n = 2 superlattice is intermediate between the short-period and the long-period (n $\ge$ 3) superlattices. Here, on one hand, the leakage of electrons to the SMO side is small enough  that the G-type AFM is maintained there as in the bulk. On the other hand the number of e$_g$ electrons leaving the LMO side is large enough that the LMO part behaves like a hole doped bulk (La$_{1-x}$Sr$_x$MnO$_3$) thereby stabilizing the FM structure as in the short period superlattice (n = 1). However, as the calculated ferromagnetic stabilization energy is relatively small here as compared to the n=1 case, it is only weakly ferromagnetic (J$_4$ = - 4 meV, Table I).

   In contrast to this, in the long-period superlattices (n $\ge$ 3), a much stronger potential barrier prevents any significant leakage of the electrons to the SMO side, except to the very first interfacial layer. This leads to the bulk magnetic behavior  inside the LMO as well as the  SMO parts. The only layers affected by the electron leakage are just two layers at the interface so that the magnetic structure as indicated in Fig. \ref{potential} is of the type ...$|$FGGF$|$FAAAF$|$..., where the vertical line indicates the interface. The ground state magnetic structures for the various superlattices discussed in this paper are consistent with those observed in the experiments\cite{anand,adamo}. 
   
\vspace{0.4cm}
{\it Electronic structure of the (LMO)$_6$/(SMO)$_3$ superlattice --}
We now turn to the electronic structure of the (LMO)$_6$/(SMO)$_3$ superlattice, which would be typical of the long-period superlattices
($n \ge 3$).
 The spin-resolved layer projected Mn-d DOS for this case is shown in Fig. \ref{dosl6s3}. 
 In a solid with complex magnetic structure, it is convenient to discuss
 the electron occupancy with a local spin quantization axis  defined with respect to the local moment of a specific magnetic atom. This was done in Fig. \ref{dosl6s3}.

 As seen from the figure, deep inside both the SMO and LMO parts, the 
 electron occupancies are more or less similar to those of the respective
 bulk compounds. The bulk behavior occurs already beyond just one Mn layer on either side of the interface. In the SMO part (Mn-1 densities, topmost panel), the Mn-t$_{2g}$ spin-up states are filled
 while the e$_g$ states are empty just like bulk SMO. In the LMO part (Mn-2 densities, bottommost panel), the e$_g$ states are
 Jahn-Teller split into two bands, with the lower one occupied, again, as in the bulk LMO.\cite{sashilmo}
 As one approaches the interface from the LMO side, the e$_g$ occupancy is reduced slightly from one due to the leakage of the 
 electron to the interfacial Mn0 layer. The transferred e$_g$ electron across the interface
 controls the magnetic behavior of the interfacial layers as already discussed.

  In summary, we have studied the change in the magnetic properties of the (LMO)$_{2n}$/(SMO)$_n$ superlattices as a function of the layer thickness `n' and explained the observed magnetic structure in terms of the electronic structure and the electron leakage across the interface. For the short period superlattice (n = 1), we find a weak variation of the potential leading to the spreading of the Mn-e$_g$ electrons throughout the superlattice, resulting in a FM structure via the carrier-mediated Zener double exchange, much like the alloy compound La$_{2/3}$Sr$_{1/3}$MnO$_3$.  For higher `n' there is a potential barrier restricting the electron leakage to the SMO side. For 
  n $\ge$ 3, the charge leakage is restricted to just two layers at the interface, beyond which a bulk-like electronic and magnetic structure results.

  This work was supported by the U.S. Department of Energy under Grant No. DE-FG02-00ER45818. We thank J. W. Freeland for stimulating this work and for valuable discussions.

\end{document}